\documentclass[12pt]{article}
\usepackage{graphicx}
\usepackage{epsfig}

\newcommand{\spone}{0.9}  

\newcommand{\sptwo}{1.4}
\newcommand{\spthree}{2.4}

\newcommand{\singlespace}{\edef\baselinestretch{\spone}\Large\normalsize}

\newcommand{\doublespace}{\edef\baselinestretch{\sptwo}\Large\normalsize}
\newcommand{\threespace}{\edef\baselinestretch{\spthree}\Large\normalsize}

\singlespace
\threespace
\doublespace

\everymath={\displaystyle}
\begin{document}
\doublespace

\begin{center}
{\bf Rephasing Invariance and Hierarchy
of the CKM Matrix\\
$~$\\
T.K. Kuo$^*$ and Lu-Xin Liu$^\dagger$\\
Physics Department, Purdue University, W. Lafayette, IN 47907}
\end{center}
\vspace{30pt}

\begin{center}
{\it Abstract}
\end{center}

We identified a set of four rephasing invariant parameters of the CKM 
matrix. They are found to exhibit hierarchies in powers of $\lambda^2$,
from $\lambda^2$ to $\lambda^8$.  It is shown that, at the present level
of accuracy, only the first three parameters are needed to fit all
available data on flavor physics.
\vspace{60pt}

\noindent
$^*$E-mail: tkkuo@physics.purdue.edu

\noindent
$^\dagger$E-mail: liul@physics.purdue.edu

\pagebreak

In the standard model, the esential input in understanding flavor
physics and CP violation is given by the Cabibbo-Kobayashi-Maskawa
(CKM) matrix, $V_{CKM}$, or rather, by the rephasing invariant
portions thereof, since one can always change the phases (rephasing)
of quark fields without affecting the physics.  What is remarkable
is that, to this date, all observations, both CP-conserving and
CP-violating, while greatly over-constraining the parameters in
$V_{CKM}$, are nevertheless completely in accord with this picture.
Another remarkable feature of $V_{CKM}$ is its hierarchical structure,
characterized by the Wolfenstein parameter [1],
$\lambda = |V_{us}|\cong 0.22$.  This implies that flavor processes can
be naturally classified by their strengths in powers of $\lambda$
(e.g. CP-violating phenomena only appear at
$O(\lambda^6)$ or higher).  In this paper we make use of a
manifestly rephasing invariant parametrization introduced in a
previous work [2].  It consists of six parameters,
$(x_i,y_j), i,j =1,2,3$, which satisfy two constraints, so that
any four of them can be used as a complete set.  Using known measurements,
we find that three of those $(y_2,y_1,x_2)$ are of order 
$O(\lambda^2), O(\lambda^4)$ and $O(\lambda^6)$, respectively.
There is also considerable evidence that a fourth, $y_3$,
is of order $O(\lambda^8)$, and is consistent with zero, at the
present level of accuracy in the determination of $V_{CKM}$.  It is then
shown that, for $V_{CKM}$, the set of three parameters
$(y_2,y_1,x_2)$, with $y_3=0$, suffices to describe all extant
data on flavor physics.  An accurate assessment of the value of
$y_3$ can only be done after more precise measurements become
available.

We begin by summarizing the main properties of the $(x,y)$ parametrization
and detailing its relations with other parametrizations which are
in common use.

We assume, without loss of generosity, that
$$ det~V_{CKM} = + 1.
  \eqno{(1)}
$$

\noindent
There are then six rephasing invariants defined by
$$  \Gamma_{ijk} = V_{1i} V_{2j} V_{3k},
  \eqno{(2)}
$$

\noindent
where $(i,j,k)$ is a permutation of (1,2,3).  It was proved
that all six $\Gamma_{ijk}$'s have the same imaginary part, $-iJ$,
where $J$ is the invariant CP measure [3], so that
$$  \Gamma_{ijk} = Re\Gamma_{ijk} - iJ.
  \eqno{(3)}
$$
It is useful to separate the even and odd permutation $\Gamma$'s and
define
$$ Re(\Gamma_{123}, \Gamma_{231}, \Gamma_{312}) = (x_1,x_2,x_3);
   \eqno{(4)}
$$
$$ Re(\Gamma_{132}, \Gamma_{213}, \Gamma_{321}) = (y_1,y_2,y_3).
   \eqno{(5)}
$$
They are found to satisfy two constraints,
$$ (x_1 + x_2 + x_3) - (y_1 + y_2 + y_3) = 1,
  \eqno{(6)}
$$
$$ x_1x_2 + x_2x_3 + x_3x_1 = y_1y_2 + y_2y_3 + y_3y_1 .
   \eqno{(7)}
$$
Thus, any four of the set $(x_i,y_j)$ may be used as a complete set of
parameters of $V_{CKM}$.  In addition, it was also established that
$$ J^2 = x_1x_2 x_3 - y_1y_2y_3.
  \eqno{(8)}
$$
All of the parameters $(x_i,y_j)$ take values between
$\pm 1, -1 \leq (x_i,y_j) \leq +1$, with $y_j \leq x_i$, for
any $(i,j)$.

We now turn to the relations between the set $(x,y)$ and other familiar
parametrizations.  The simplest is that with $|V_{ij}|^2$ [4],
the absolute square of the elements of $V_{CKM}$.  It is given by
$$  W = \left( \begin{array}{ccc}
     |V_{11}|^2 & |V_{12}|^2 & |V_{13}|^2\\
     |V_{21}|^2 & |V_{22}|^2 & |V_{23}|^2\\
     |V_{31}|^2 & |V_{32}|^2 & |V_{33}|^2\\
  \end{array} \right)
 \eqno{(9)}
$$

$$  = \left( \begin{array}{ccc}
     x_1-y_1 & x_2-y_2 & x_3-y_3\\  
     x_3-y_2 & x_1-y_3 & x_2-y_1\\
     x_2-y_3 & x_3-y_1 & x_1-y_2
   \end{array} \right)
  \eqno{(10)}
$$
It is also straight-forward to obtain the relations between
$(x,y)$ and the ``standard" parametrization of the Particle Data Group [5]
$$  V^{(s)} = \left( \begin{array}{ccc}
    c_{12}c_{13}   & s_{12} c_{13}   & s_{13} e^{-i\delta}\\
    -s_{12}c_{23}-c_{12}s_{23}s_{13}e^{i\delta} &
       c_{12}c_{23}-s_{12}s_{23}s_{13}e^{i\delta} &
         s_{23} c_{13} \\
    s_{12}s_{23} -c_{12}c_{23}s_{13}e^{i\delta} &
       -c_{12}s_{23}-s_{12}c_{23}s_{13}e^{i\delta} &
           c_{23}c_{13} \end{array} \right)  .
\eqno{(11)}
$$

\noindent
Note that $det~V^{(s)} = +1$.  However, $V^{(s)}$ is not invariant
under rephasing.  We will not write down the explicit relations between
$(x,y)$ and the angles in $V^{(s)}$.  These relations are exact but they
tend to be rather cumbersome.  

Another well-established parametrization,
due to Wolfenstein [1], is often used.  It is an expansion of
$V_{CKM}$ in the parameter $\lambda = |V_{us}| \cong 0.22$,
whose validity is related to the fact that the three angles in 
$V^{(s)}$ are small and hierarchical.
Wolfenstein's representation of $V_{CKM}$ reads:
$$  V^{(W)} = \left( \begin{array}{ccc}
     1-\lambda^2/2 & \lambda & A\lambda^3(\rho-i\eta) \\
     -\lambda & 1-\lambda^2/2 & A\lambda^2 \\
     A\lambda^3(1-\rho-i\eta) & -A\lambda^2 & 1 \end{array}\right)
    + O(\lambda^4).
  \eqno{(12)}
$$
Note that $det~V^{(W)} = 1 + O(\lambda^2)$, and, like $V^{(s)}$,
$V^{(W)}$ is not rephasing invariant.  A systematic expansion
into higher orders of $\lambda$ is also available [6],
and is often used in the literature.  In the spirit of an expansion
in $\lambda$, we may compute $(x,y)$ by using $\Gamma_{ijk}^{(W)} =
V_{1i}^{(W)} V_{2j}^{(W)} V_{3k}^{(W)}$, which form rephasing
invariant combinations.  We find, to leading order in $\lambda^2$,
$$  \Gamma_{123}^{(W)} = 1 - \lambda^2 = x_1,
  \eqno{(13)}
$$
$$  \Gamma_{213}^{(W)} = -\lambda^2 = y_2,
   \eqno{(14)}
$$
$$ \Gamma_{132}^{(W)} = -A^2\lambda^4 = y_1.
  \eqno{(15)}
$$
None of these $\Gamma$'s contains the imaginary part,
$-iJ$, which is $O(\lambda^6)$,  
although they do give correct values for 
$x_1,y_2$ and $y_1$, to leading order.  On the other
hand,
$$  \Gamma_{231}^{(W)} = A^2\lambda^6 [(1-\rho)-i\eta] = x_2 -iJ,
   \eqno{(16)}
$$
$$  \Gamma_{321}^{(W)} = A^2\lambda^6 [(\rho-\rho^2-\eta^2)-i\eta]
       = y_3-iJ,
   \eqno{(17)}
$$
$$\Gamma_{312}^{(W)} = A^2\lambda^6(\rho-i\eta) = x_3 -iJ.
  \eqno{(18)}
$$

\noindent
All of these quantities are $O(\lambda^6)$, and they have the
same imaginary part, with 
$$  J = A^2\lambda^6 \eta,
   \eqno{(19)}
$$
which is a well-known approximate result.  Had we used the improved
version of $V^{(W)}$ [6], it will be seen that, except for
$\Gamma_{123}^{(W)}$, all $\Gamma^{(W)}$'s contain the same
imaginary part, $-iJ$.  

The above analysis shows that, to leading order in $\lambda^2$,
the $(x,y)$ parameters are given in the hierarchical order
$x_1 = O(1)$, $x_2 = O(\lambda^6)$, $x_3 = O(\lambda^6)$, $y_1 = 
O(\lambda^4)$, $y_2 = O(\lambda^2)$ and $y_3 = O(\lambda^6)$. These 
results are obtained with the tacit assumption that, in the 
parametrization $V^{(W)}$,
the values $(A,\rho,\eta)$ are all of order one and are unrelated
to each other.  When we incorporate recent measurements, it turns
out that there are correlations and detailed structures which have
interesting implications.  This we will discuss in the following.

We begin by improving upon our earlier analysis in Ref. [2], and
write down more precise relations between the $(x,y)$ parameters
and the angles $(\alpha,\beta,\gamma)$ of the unitarity triangle.
From the unitarity condition, 
$$  V_{11}V_{13}^* + V_{21}V_{23}^* + V_{31}V_{33}^* = 0,
  \eqno{(20)}
$$
we obtain a rephasing invariant equation by multiplying by
$V_{11}^*V_{13}$.  The result is a triangle in the complex plane
whose height is (exactly) $iJ$, with one side given by
$$V_{11}^*V_{13}V_{31}V_{32}^* = y_3 + |V_{13}|^2|V_{31}|^2 - iJ.
  \eqno{(21)}
$$
This triangle is plotted in Fig.1.

Since $|V_{13}|^2 |V_{31}|^2 \leq O(\lambda^{12})$, together with
the estimate (to be discussed later) $ y\approx O(\lambda^8)$,
we have, from Fig. 1,

\begin{figure}
\centering
\includegraphics[width=6cm,height=6cm]{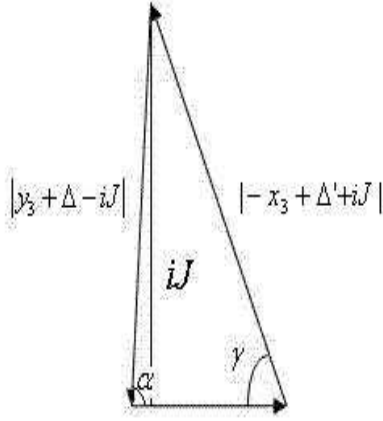}
\caption{Rescaled unitarity triangle with base 
$|V_{11}|^2 |V_{13}|^2$ and sides as labelled. 
Also, $\Delta=|V_{13}|^2 |V_{31}|^2=O(\lambda^{12}), 
\Delta '=|V_{13}|^2 |V_{21}|^2=O(\lambda^8).$}\label{fig:secondgraph3.eps} 
\end{figure}

$$  \tan \alpha =-(J/y_3) (1+ O(\lambda^4)).
  \eqno{(22)}
$$
Similarly, using Ref.[2], we find
$$  \tan\beta = (J/x_2) (1+ O(\lambda^4)),
  \eqno{(23)}
$$
$$\tan\gamma = (J/x_3) (1 + O(\lambda^2)).
 \eqno{(24)}
$$

\noindent
Thus, to a high degree of accuracy, the angles $(\alpha,\beta,\gamma)$
are the phase angles of $(\Gamma_{321}, \Gamma_{231}^*, \Gamma_{312}^*)$.
(Note that, in Ref.[2], Eq.(60) has a wrong sign, and the statement 
thereafter contains a typo.)

Experimentally, $\beta$ has been rather precisely measured,
while $\alpha$ and $\gamma$ are not so well determined.
To assess their effects on the $(x,y)$ parameters, we make
use of currently available global fits [7,8].  Because the
two analyses yield similar results, we will only quote the
values from Ref.[7], for simplicity.
The values of $(\alpha,\beta,\gamma)$, in degrees, are given by
$$\alpha = 94_{-10}^{+12},~~ \beta = 24 \pm 2, ~~
   \gamma = 62_{-12}^{+10}.
  \eqno{(25)}
$$
With these we find

$$ x_2/J = 2.24 \pm 0.2,
 \eqno{(26)}
$$
$$ x_2/x_3 = 4.3_{-1.3}^{+2.7},
  \eqno{(27)}
$$
$$ y_3/x_2 = 0.03_{-0.07}^{+0.1},
 \eqno{(28)}
$$
and, using $\gamma = \pi -(\alpha+\beta)$ to correlate
$\gamma$ and $\alpha$,
$$ y_3/x_3 = 0.13_{-0.4}^{+0.2}.
 \eqno{(29)}
$$
These results suffer from large uncertainties.  However, it is
clear that the parameters $x_2, J, x_3, y_3$, are not of the same
order. In particular, we find
$$ y_3/x_2 = O(\lambda^2).
  \eqno{(30)}
$$
In fact, the result $y_3/x_2 = O(\lambda^2)$ can be expressed
in a variety of forms when we write $y_3 = Re\Gamma_{321}$ in
different parametrizations.  

In the Wolfenstein parametrization,
$$ Re\Gamma_{321} = A^2 \lambda^6 [\rho -(\rho^2+\eta^2)].
 \eqno{(31)}
$$
For $V^{(s)}$, we have
$$ \begin{array}{rcl} 
   Re\Gamma_{321} &=& Re[s_{13}(c_{12}c_{23}-s_{12}s_{23}
    s_{13}e^{i\delta})(e^{-i\delta} s_{12}s_{23}
    -c_{12}c_{23}s_{13})]\\
   &\cong & s_{13} (c_\delta s_{12}s_{23}-s_{13}).
  \end{array}
 \eqno{(32)}
$$
Using Eq.(10), $|V_{ub}|^2 = x_3-y_3$,
$|V_{td}|^2 = x_2-y_3$, it follows that
$$  Re\Gamma_{321} = \frac{1}{2} [-|V_{td}|^2 -
    |V_{ub}|^2 + |V_{us}|^2 |V_{cb}|^2],
 \eqno{(33)}
$$
where we have used the constraint (Eq.(7))
$(x_2 + x_3 \cong |V_{us}|^2 |V_{cb}|^2$, with $x_1 \cong 1)$.
Thus, the condition $y_3/x_2 = O(\lambda^2)$ takes on a variety
of forms:
$$ 1)~~~-\tan \beta/\tan \alpha = O(\lambda^2);
  \eqno{(34)}
$$
$$ 2)~~~[\rho - (\rho^2+\eta^2)]/(1-\rho) = O(\lambda^2);
  \eqno{(35)}
$$
$$ 3)~~~\cot^2\delta (c_\delta s_{12}s_{23} -s_{13})/s_{13}
      = O(\lambda^2);
  \eqno{(36)}
$$
$$ 4)~~~\frac{1}{2}(|V_{us}|^2 |V_{cb}|^2 - |V_{td}|^2 -
     |V_{ub}|^2)/|V_{td}|^2 = O(\lambda^2) .
  \eqno{(37)}
$$
Here, in 3), we have used the approximate relations
$\delta = \gamma$, $\tan\beta \tan\gamma = 1$ to obtain
$x_2 = x_3 \tan^2\delta  =  s_{13}^2 \tan^2 \delta$.
All of these relations are reasonably well satisfied by
using the values of the global CKM fits [7].  They are,
approximately, $\rho=0.19 \pm 0.08, \eta=0.36 \pm 0.05; s_{12}=0.226 \pm 
0.002$, $s_{13} = (3.9 \pm 0.3) \times 10^{-3},
s_{23} = (41 \pm 1) \times 10^{-3}$,
$\delta = 62 \pm 10$; $|V_{us}| = 0.23 \pm 0.002$,
$|V_{ub}| = (3.9 \pm 0.3)\times 10^{-3}$, 
$|V_{cb}| = (41 \pm 1)\times 10^{-3}$, $|V_{td}| =
(8.3 \pm 0.8) \times 10^{-3}$.  Eqs.(34-37) exhibit the
intriguing correlations amongst the $V_{CKM}$ parameters.
Without them $y_3$ would be of the same order as $x_2$.

The above analysis can be summarized as an expansion in
$\lambda^2$:
$$ (-y_2, -y_1, x_2, y_3) = (\lambda^2,A^2\lambda^4, B^2\lambda^6,
   C\lambda^8),
  \eqno{(38)}
$$
with
$$ A^2 = 0.64 \pm 0.05
  \eqno{(39)}
$$
$$ B^2 = 0.5 \pm 0.1.
  \eqno{(40)} 
$$
The parameter $C$ is very poorly determined.  We use the above
relations and make a rough estimate to find
$$ C = 0.3 \pm 1.
  \eqno{(41)}
$$
Thus, a convenient parametrization of $V_{CKM}$ is to use the
set $(y_2,y_1,x_2,y_3)$.  It gives rise to an expansion in powers
of $\lambda^2$, in contrast to the use of $V^{(W)}$ or
$V^{(s)}$, whose matrix elements can be regarded as expansions
in powers of $\lambda$.  The difference originates from rephasing
considerations.  For, when one uses rephasing invariants (which 
are what enter into physical quantities) constructed
out of $V^{(W)}$ or $V^{(s)}$, such as $|V_{ij}|^2$ or 
$V_{\alpha i} V_{\beta j} V_{\alpha j}^* V_{\beta i}^*$ or
$(x_i,y_j)$, it is seen that the expansion parameters is
$\lambda^2$, and not $\lambda$. Thus, e.g., $|V_{ij}|^2$ have values which
are of order $O(1),O(\lambda^2),O(\lambda^4),O(\lambda^6)$. However, the 
combination $y_3$ (Eq.(33)) contains a cancellation, resulting in 
$y_3=O(\lambda^8)$. It is this feature which distinguishes the $(x,y)$ set
from other parametrizations.

The result $y_3 = O(\lambda^8)$ has another interesting consequence.
To the extent that all available measurements on $V_{CKM}$ are only
accurate to $O(\lambda^6)$, we should be able to set 
$y_3 = 0$ and fit the extant data on $V_{CKM}$ by three
parameters, $(y_2,y_1,x_2)$.  This will be presented in Fig.2.
The inputs are taken from those of the well-known unitarity triangle
construction [7,8,9] in terms of $\rho$ and $\eta$ (since we are
only interested in leading order effects, we take
$\bar{\rho} = \rho, \bar{\eta} = \eta)$.  With
$y_2 = -A^2 \lambda^4, y_1 = -\lambda^2, y_3 = 0, x_1 =1 -\lambda^2$,
all of the physical quantities $(\epsilon_K,|V_{ub}|^2,
|V_{td}|^2, \tan \beta)$ can be converted from functions of
$(\rho,\eta)$ into those of $(x_2,x_3)$ using
$A^2 \lambda^6 (1-2\rho) = x_2 -x_3$,
$A^2\lambda^6 \eta = \sqrt{x_2x_3}$.  In addition, we use
$J^2 = x_2x_3$ and the constraint $x_2 + x_3 \cong y_1y_2$.
The values of the physical quantities are taken from Ref.[9].
It is seen that there is a shaded region in the $(x_2,x_3)$ plane where
all constraints meet.  Fig.2 represents a three parameter
$(y_2,y_1,x_2)$ fit (with $x_3 = y_1y_2 -x_2)$ to existing data
on $V_{CKM}$.  Qualitatively, we can understand the viability of a 
three parameter fit as follows. The parameter $\eta$, as determined 
in Ref.[7], has errors of order $O(\lambda^2)$. Setting $y_3=0$, 
according to Eq.(35), amounts to eliminating $\eta$ and using 
$\eta=\sqrt{\rho-\rho^2}$. This relation is satisfied to 
$O(\lambda^2)$. Thus, Fig.2 represents a fit with the parameter  
set $(\lambda^2,A^2 \lambda^4, \rho,\eta=\sqrt{\rho-\rho^2})$, and 
should be robust at the $O(\lambda^2)$ level, as demonstrated. 
Similar results also follow if we use other parametrizations. The above 
analysis also suggests that whether a non-vanishing $y_3$ is needed has 
to wait until more precision measurements become available. 

\begin{figure}[htp]
\centering
\includegraphics[width=12cm]{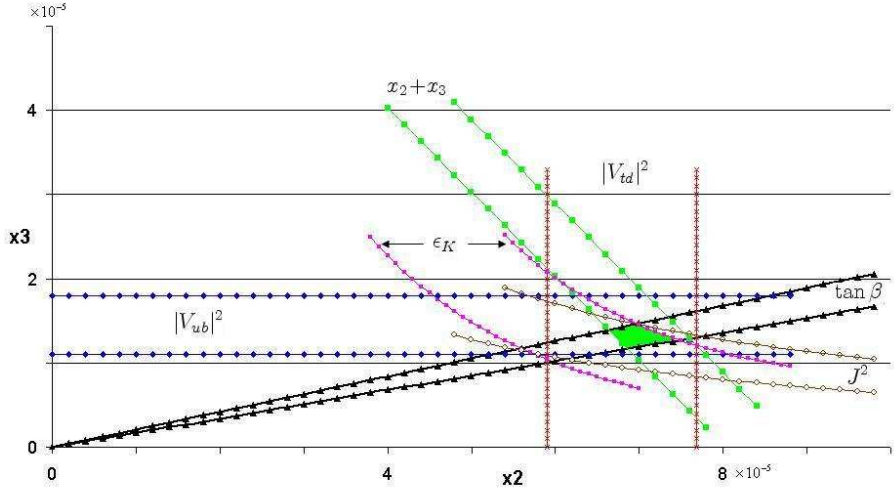}
\caption{A three parameter fit to $V_{CKM}$, 
with fixed $y_1(=-A^2 \lambda^4), 
y_2(=-\lambda^2)$, and the constraint 
$x_2+x_3=y_1 y_2$. Values of the physical 
quantities are taken from Ref.[9].}\label{fig:ckm-graph12.eps} 
\end{figure}

We close with a few concluding remarks.  In this paper we argue that,
although our present knowledge on $V_{CKM}$ is still far from
precise, it nevertheless offers considerable evidence of intriguing
correlations, as in Eqs.(34-37).  We can take advantage of this
by using a particular set of parameters, $(y_2,y_1,x_2,y_3)$,
which gives rise to a natural expansion in powers of $\lambda^2$.
It is shown that, at the current level of accuracy, we can set
$y_3 = O(\lambda^8) \cong 0$ and obtain a three parameter fit to 
$V_{CKM}$.
Within the currently available data set, the proper way to estimate
a ``best" value for $y_3$ would be to do a global fit with the
$(x,y)$ variables.  However, this is beyond the scope of the present
paper.  Of course, we would have a better handle in the future,
when more and better measurements are performed.

One important motivation in analyzing $V_{CKM}$ is to check its
consistency by over-contraining its parameters, with an eye for
discrepancies which might originate from ``new physics".  It is
our hope that a parametrization which has a distinct hierarchy
can help to identify features that will stand out, so that the
success or failure of a model can be better assessed.

\begin{center}
This work is supported in part by DOE grant No.
DE-FG02-91ER40681.
\end{center}

\pagebreak

\begin{center}
{\bf REFERENCES}
\end{center}

\begin{description}
\item[[1]] L. Wolfenstein, Phys. Rev. Lett. {\bf 51}, 1945 (1983).

\item[[2]] T.K. Kuo and Tae-Hun Lee, Phys. Rev. {\bf D71}, 093011 (2005).

\item[[3]] C. Jarlskog, Phys. Rev. Lett. {\bf 55}, 1039 (1985); also in $CP$
Violation, Edited by C. Jarlskog (World Scientific, Singapore, 1989).

\item[[4]] C. Hamzaoui, Phys. Rev. Lett. {\bf 61}, 35 (1988);
G.C. Branco and L. Lavoura, Phys. Lett. {\bf B208}, 123 (1988).

\item[[5]] Particle Data Group, Phys. Lett. {\bf B952}, 130 (2004);
L.L. Chau and W.Y. Keung, Phys. Rev. Lett. {\bf 53}, 1802
(1984); H. Harari and M. Leurer, Phyus. Lett. {\bf B181}, 123 (1986).

\item[[6]] A.J. Buras, M.E. Lautenbacher and G. Ostermaier, Phys. Rev. {\bf D50},
3433 (1994).  

\item[[7]] J. Charles et al. [CKMfitter Collaboration], hep-ph/0406184.

\item[[8]] M. Bona et al. [UTfit Collaboration], hep-ph/0501199.

\item[[9]] See, e.g., A.J. Buras, hep-ph/0505175, and references therein.

\end{description}

\end{document}